# Submillimeter-Accurate 3D Lumbar Spine Reconstruction from Biplanar X-Ray Images: Incorporating a Multi-Task Network and Landmark-Weighted Loss

Wanxin Yu [1,2], Zhemin Zhu [4], Cong Wang [1,2], Yihang Bao [1,2], Chunjie Xia[5], Rongshan Cheng [1,2]*, Yan Yu [3]*, Tsung-Yuan Tsai [1,2,4]*

[1] *School of Biomedical Engineering & Med-X Research Institute, Shanghai Jiao Tong University, Shanghai 200030, China.*

[2] *Engineering Research Center of Digital Medicine and Clinical Translation, Ministry of Education, Shanghai 200030, China.*

[3] *Department of Spine Surgery, Tongji Hospital, School of Medicine, Tongji University, Shanghai, 200065, China.*

[4] *TAOiMAGE Medical Technologies Corporation, Shanghai 200120, China.*

[5] *School of Mechanical Engineering, Shanghai Jiao Tong University, Shanghai 200030, China.*

***Corresponding Author:**

**Rongshan Cheng,** *School of Biomedical Engineering & Med-X Research Institute, Shanghai Jiao Tong University, Shanghai 200030, China. Email: chengrongshan@sjtu.edu.cn*

**Yan Yu,** *Department of Spine Surgery, Tongji Hospital, School of Medicine, Tongji University, Shanghai, 200065, China. E-mail: yyu15@tongji.edu.cn*

**Tsung-Yuan Tsai,** *School of Biomedical Engineering & Med-X Research Institute, Shanghai Jiao Tong University, Shanghai, 200030, China. Email: tytsai@sjtu.edu.cn*

**Abstract:**
To meet the clinical demand for accurate 3D lumbar spine assessment in a weight-bearing position, this study presents a novel, fully automatic framework for high-precision 3D reconstruction from biplanar X-ray images, overcoming the limitations of existing methods. The core of this method involves a novel multi-task deep learning network that simultaneously performs lumbar decomposition and landmark detection on the original biplanar radiographs. The decomposition effectively eliminates interference from surrounding tissues, simplifying subsequent image registration, while the landmark detection provides an initial pose estimation for the Statistical Shape Model (SSM), enhancing the efficiency and robustness of the registration process. Building on this, we introduce a landmark-weighted 2D-3D registration strategy. By assigning higher weights to complex posterior structures like the transverse and spinous processes during optimization, this strategy significantly enhances the reconstruction accuracy of the posterior arch. Our method was validated against a gold standard derived from registering CT segmentations to the biplanar X-rays. It sets a new benchmark by achieving sub-millimeter accuracy and completes the full reconstruction and measurement workflow in under 20 seconds, establishing a state-of-the-art combination of precision and speed. This fast and low-dose pipeline provides a powerful automated tool for diagnosing lumbar conditions such as spondylolisthesis and scoliosis in their functional, weight-bearing state.

## 1 Introduction

The lumbar spine is the most heavily loaded part of the spine, and its pathological conditions are often accompanied by changes in three-dimensional (3D) morphology and structure. Therefore, understanding the 3D alignment of the lumbar spine and the morphology of the vertebral bodies is crucial for the clinical diagnosis of spinal diseases. For instance, the height and endplate area of the vertebral bodies can help detect vertebral fractures, collapse, or deformities [1,2]. Abnormal intervertebral disc space may indicate issues such as intervertebral disc degeneration, herniation, or protrusion [3]. Moreover, understanding the 3D curvature of the spine can be beneficial for diagnosing scoliosis and for surgical planning [4]. The 3D structure of the lumbar spine in the standing position can better reveal the anterior displacement of the vertebral bodies, aiding in the detection of degenerative lumbar spondylolisthesis [5]. The patient-specific 3D anatomy of the lumbar spine can help measure vertebral geometry and pathological deformities, assisting in disease assessment, surgical planning, and implant design[6]. The 3D parameters of the weight-bearing lumbar spine hold greater clinical value for diagnosis and surgical planning [7,8]. However, there is currently a lack of automated imaging devices or methods to establish the 3D structure of the patient's weight-bearing spine. Traditional X-rays only provide two-dimensional (2D) information and are prone to issues such as anatomical structure overlap and projection distortion. The anatomical parameters measured from X-rays are insufficient to accurately reflect the true condition [9,10]. Conventional 3D medical imaging techniques, such as CT and MRI, primarily depict 3D morphology in the supine position [11], and have drawbacks including high radiation exposure, high costs, lengthy imaging times, and the necessity for manual segmentation and reconstruction [12]. These traditional medical imaging methods require physicians to perform clinical parameter measurements, which can be time-consuming, laborious, and exhibit poor reproducibility. Consequently, fully automated 3D reconstruction of the lumbar spine based on biplanar orthogonal X-ray images has emerged as a low-cost and safe solution for obtaining 3D parameters of the lumbar spine in the standing position. This method can be applied to evaluate lumbar scoliosis and spondylolisthesis under load, as well as for repetitive follow-up imaging and pediatric assessments [13].

Currently, there are two main approaches for reconstructing the 3D bone structure from biplanar X-ray images: end-to-end reconstruction based on machine learning and registration-based reconstruction using deformable models. The former approach suffers from low reconstruction accuracy (>3mm) due to the lack of prior shape information about the reconstructed structure and the limited information provided by the biplanar images [14]. The latter approach begins with encoding the prior appearance information of the anatomical structure [15]. It then uses a set of parameters to manipulate the model and adjust the deformable model parameters based on the patient's X-ray images, achieving high-precision patient-specific 3D reconstruction [12,16]. There are two main strategies for finding the optimal parameters in this approach. The first involves identifying landmarks in the X-ray images, establishing correspondence between the deformable model and the landmarks, and minimizing the geometric distance between them [17,18]. The methods for identifying landmarks range from fully manual to fully automatic; however, the manual annotation method is very time-consuming. Conversely, fully automatic methods tend to have larger errors and are more prone to mistakes [12,19]. The second method is based on digitally reconstructed radiograph (DRR), which projects the deformable model as a whole onto the 2D plane of the X-ray image to obtain a simulated X-ray image. This method iteratively improves the similarity between the raw X-ray image and the DRR to obtain best-fitting parameters for the deformable model [12,20–22]. The DRR-based method is more accurate than the feature-based method, but it takes several minutes to compute and is more prone to getting stuck in local optima [15,23].

As a technology used to assist clinical diagnosis, the primary goal of 3D reconstruction of human anatomical structures based on simultaneous acquisition of anteroposterior (AP) and lateral (LAT) radiographic images is to achieve higher accuracy. Taking the lumbar spine as an example, the existing publicly available methods based on variable models can achieve an accuracy of around 1.0–2.0 mm, as shown in Table 1I. However, such effects are still insufficient to meet clinical needs. Measurement errors for vertebral height and disc thickness should be within 1 mm, and diagnostic precision for spinal canal stenosis must reach submillimeter levels to avoid missed or incorrect diagnoses[24,25]. Besides, current methods often overlook detailed anatomical structures, such as the transverse and spinous processes of the lumbar spine. This can result in models that emphasize the larger vertebral bodies during the 2D-3D registration process while neglecting the more complex structures of the vertebral arches. Most importantly, the ground truth used by most studies to evaluate the 3D reconstruction results in current research cannot be regarded as the gold standard. Some studies utilize CT/MRI segmentation results as the validation standard, which cannot verify the 3D positioning of the spine [16,26] Other studies employ semi-automatic reconstructions with manually adjusted shapes as the validation standard, which fail to confirm the 3D shape information of the vertebra [18,22].

In summary, current 3D lumbar spine reconstruction from biplanar X-rays faces challenges related to insufficient accuracy to meet clinical requirements and a lack of focus on complex vertebral arch structures[27,28]. This study aims to improve the 3D reconstruction accuracy of the lumbar spine to sub-millimeter levels by pre- preprocessing X-rays with lumbar decomposition and landmark detection, and employing a landmark-weighted approach to complete the 2D-3D registration of the deformable model. The outcomes will be validated against CT segmentation and biplanar X-ray registration results as the gold standard.

## 2 Proposed method

The overall workflow of our proposed method is illustrated in Fig. 1. First, we constructed Statistical Shape Models (SSMs) of the L1-L5 vertebrae, which serve as a prior for the 3D vertebral shape and constrain their deformation during registration. Second, we performed lumbar decomposition and vertebral landmark detection on the raw biplanar X-ray images. The so-called 'decomposition' involves segmenting the lumbar region from the raw images and rendering it in the style of a DRR. This approach eliminates interference from other tissues and simplifies the image similarity metric for the subsequent registration process. Next, a coarse alignment of the SSMs was performed based on the predicted vertebral landmark positions and the projection geometry. Finally, the definitive 3D lumbar reconstruction was achieved by iteratively optimizing the similarity between the DRRs generated from the models and the decomposed lumbar images.

### 2.1 A. Statistical Modeling

The pipeline for constructing the SSMs is depicted in Fig. 2. The process began by adaptively re-meshing individual vertebrae to increase vertex density in regions with greater geometric variation. The mesh with the minimum cumulative distance to all other meshes in the dataset was selected as the reference mesh. Subsequently, rigid transformations were applied to each target mesh for alignment with the reference mesh. Initial rigid alignment was achieved by establishing local vertebral coordinate systems using bony landmarks and transforming vertex coordinates accordingly. This alignment was further refined with an optimized Iterative Closest Point (ICP) algorithm to exclude outlier correspondences, followed by Generalized Procrustes Analysis (GPA) without scaling. This non-scaling approach preserves vertebral size as an important mode in the subsequent Principal Component Analysis (PCA). Next, non-rigid alignment was performed by approximating the vertex displacement between the reference and target meshes using Gaussian radial basis functions with thin-plate spline interpolation [29]. In the final step, PCA was applied to the collection of aligned shape coordinates to compute the primary modes of variation and establish the mean shape for each vertebral level. At last, four anatomical landmarks (endplate centers and transverse process tips) and the superior and inferior endplates on each SSM were annotated by the expert clinician to guide the pre-registration for subsequent 3D reconstruction and enable the automatic measurement of lumbar parameters.

## 2.2 B. X-ray Image Preprocessing

The overall architecture of the multitask network for lumbar decomposition and landmark detection proposed in this paper is shown in Fig. 3. The encoder portion of the convolutional neural network employed shared parameters to conserve computational resources. The decoder was divided into two paths: one for lumbar decomposition and the other for vertebral landmark detection. Additionally, a feature fusion module (FFM) was implemented to combine the features from both tasks. Finally, their respective output heads were utilized to finalize the results of the two tasks.

**Encoder:** Compared with widely used backbone networks such as ResNet, UNet, and DeepLab, which tend to lose detailed spatial information due to repeated downsampling, HRNet maintains high-resolution representations throughout the feature extraction process [30]. This capability allows the model to better capture spatial details and edge information, which is critical for accurate prediction in tasks with fine-grained requirements. We adopt HRNet-W48 as the encoder of the multi-task prediction network. HRNet-W48 has four stages with parallel multi-resolution branches, outputting feature maps at 1/4, 1/8, 1/16, and 1/32 of the input size ($\{F_1, F_2, F_3, F_4\}$), with channel numbers [48, 96, 192, 384]. This architecture better preserves spatial details and improves prediction accuracy compared to conventional backbones.

**Decoder:** The decoder employs a U-Net-like architecture, separating features through convolutional layers to obtain the decomposition path and the landmark path. The features in the two paths are then upsampled and progressively decoded. We posit that there are significant interdependencies between lumbar decomposition results and vertebral landmark locations. To leverage this relationship, we designed the FFM to explicitly incorporate spatial dependencies between the two decoding pathways, thereby enhancing the predictive accuracy of both tasks. Its architecture is illustrated in Fig. 4. The decomposition features and landmark features are input to the FFM, which first uses element-wise summation to fuse the information between the decomposition and landmark paths. Next, convolution operations extract the spatial dependencies from the fused feature map, effectively employing trainable parameters for image alignment. The fused feature map is then concatenated with the original feature maps of the decomposition and landmark paths to ensure that no information is lost. Convolution operations are subsequently performed to maintain consistency in the number of channels with the input feature maps. Finally, upsampling is carried out, and element-wise summation with the previous level feature map is performed to introduce higher-dimensional features, which are then input into the next stage of decoding.

**Mapping layer:** The network output was divided into the results of lumbar decomposition and the detection of seven landmarks on each lumbar vertebra. The lumbar decomposition result is a grayscale image, while the landmark detection result includes the heatmap of the vertebral centers, the center offset vector map, and the landmark offset vector map. Due to the complex X-ray structure of the lumbar region, directly obtaining the coordinates of the landmarks is challenging. Therefore, we transformed the landmark detection task into predicting the position of the center point of each vertebra and the offset vector of each landmark. This approach has the advantage of ensuring that the predicted results for the vertebral centers do not overlap. Additionally, there is a dependence between the vertebral centers and the positions of the landmarks, which enhances the accuracy of landmark detection [31].

**Loss:** The total regression loss is comprised of two parts: lumbar decomposition and landmark detection. For the lumbar decomposition component, we employed the normalized cross-correlation (NCC) loss and the $L_2$ loss as the optimization functions. The NCC loss effectively detects local features or patterns in the image, and it is defined as:

$$L_{NCC} = \frac{\sum_{i=1}^{N}\left(I_{gt}^{i} - \overline{I_{gt}}\right)\left(I_{p}^{i} - \overline{I_{p}}\right)}{\sqrt{\sum_{i=1}^{N}\left(I_{gt}^{i} - \overline{I_{gt}}\right)^{2} \cdot \sum_{i=1}^{N}\left(I_{p}^{i} - \overline{I_{p}}\right)^{2}}} \tag{1}$$

Here, N represents the total number of pixels in the image, $I_{gt}^{i}$ and $I_{p}^{i}$ represent the pixel values of the ground truth image and the predicted image respectively, and $\overline{I_{gt}}$ and $\overline{I_{p}}$ represent their respective means. Since we want the output to be the projection image of the lumbar vertebrae, we need not only to segment the position of the lumbar vertebrae but also to convert it to the DRR projection style. Therefore, we also employ a simple L2 loss:

$$L_{L2} = \frac{1}{N}\sum_{i=1}^{N}\left(I_{gt}^{i} - I_{p}^{i}\right)^{2} \tag{2}$$

The landmark detection loss includes the vertebral center heatmap loss $L_{hm}$, the center offset loss$L_{co}$, and the landmark offset loss $L_{lo}$. $L_{hm}$ is defined as:

$$L_{\mathrm{hm}} = -\sum_{i=1}^{N}\left[\mathrm{I}\left(I_{gt}^{i} = 1\right) \cdot \log\left(I_{p}^{i}\right) \cdot \left(1 - I_{p}^{i}\right)^{2} + \mathrm{I}\left(I_{gt}^{i} < 1\right) \cdot \left(1 - I_{p}^{i}\right)^{4} \cdot \log\left(1 - I_{p}^{i}\right) \cdot {I_{p}^{i}}^{2}\right] \tag{3}$$

Where $\mathrm{I}(\cdot)$ is the indicator function. $L_{\mathrm{hm}}$ can give greater weight to samples that are difficult to classify when dealing with imbalanced datasets, thereby improving the robustness of the model. $L_{co}$ and $L_{lo}$ both use a simple L1 loss:

$$L_{co} = \frac{1}{5}\sum_{i=1}^{5}\left|c_{gt}^{i} - c_{p}^{i}\right| \tag{4}$$

$$L_{lo} = \frac{1}{35}\sum_{i=1}^{35}\left|l_{gt}^{i} - l_{p}^{i}\right| \tag{5}$$

Where $c_{gt}^{i}$ and $c_{p}^{i}$ are the ground truth and predicted values of the offset coordinates of the centers of the five lumbar vertebrae, respectively, and $l_{gt}^{i}$ and $l_{p}^{i}$ are the ground truth and predicted values of the offset coordinates of the 35 feature points on the five lumbar vertebrae, relative to their corresponding vertebral centers. We introduced the weight hyperparameters $\lambda_{NCC}, \lambda_{L2}, \lambda_{hm}, \lambda_{co}, \lambda_{lo}$ to balance the scale and importance of each loss term, resulting in the following total training loss function:

$$\text{Loss} = \lambda_{NCC}L_{NCC} + \lambda_{L2}L_{L2} + \lambda_{hm}L_{hm} + \lambda_{co}L_{co} + \lambda_{lo}L_{lo} \quad (6)$$

2.3 C. Automatic 3D Reconstruction

We complete the fully automatic 3D reconstruction of the lumbar vertebrae through 2D-3D registration of the SSM and the lumbar decomposition image. To improve registration efficiency and prevent the process from getting stuck in a local optimum, we first initialize the pose of each vertebra. First, a local coordinate system was established for each SSM based on the expert-annotated centers of the superior and inferior endplates and the transverse process landmarks. Second, given the projection environment of the biplanar system and the positions of the vertebral landmarks predicted by the network, we can calculate the optimal fitting positions of these landmarks in 3D space using the least squares method:

$$\text{I} = O_1 + \hat{x}_0 \cdot (P_1 - O_1) \quad (7)$$

$$\hat{x} = \underset{x}{\text{argmin}}\left|\left|Ax - b\right|\right|^2$$

Where $\text{A} = [\begin{matrix} P_1 - O_1 \\ O_2 - P_2 \end{matrix}]$, $\text{b} = O_2 - O_1$. $O_1, O_2$ are the 3D coordinates of the radiation sources for the AP and LAT views, respectively. $P_1$ and $P_2$ are the 3D coordinates of the same landmark on the AP and LAT detectors, respectively. By calculating each lumbar vertebra's transformation matrix from the reconstructed centers of the superior and inferior endplates and the transverse process positions, we transform the vertex coordinates of the SSMs to their estimated initial positions within the biplanar projection environment.

The next step is to perform the 2D-3D registration process. For each vertebra, it is necessary to optimize its six degrees of freedom (6-DOF) rigid transformation parameters and the non-rigid transformation parameters controlled by the principal components. As shown in Fig. 5, the first 15 principal components were selected based on the analysis from Cattell's Scree Test.

We constrained the range of the parameter search based on the accuracy of pre-alignment. Translation along the three orthogonal axes was limited to ±10mm, axial rotation around the axes was limited to ±20°. The deformation of each principal component was confined within three standard deviations. During the registration process, a new shape model instance was constructed by searching for the rigid transformation and non-rigid deformation parameters of the shape model. DRRs of the generated instances were then created, with the projection parameters consistent with those used for generating vertebrae separation labels. Each time a new single lumbar vertebra instance is generated, it can be represented as:

$$\text{S} = \left(\bar{S} + \sum_{i=1}^{15} P_i\, b_i\right) \cdot T \quad (8)$$

where S represents the ordered arrangement of the vertex coordinates of the vertebral surface mesh, $\bar{S}$ denotes the mean shape model of the lumbar vertebra, $P_i$ is the eigenvector of its $i$-th principal component, $b_i$ is its corresponding score, and T is the 3D transformation matrix established according to its 6-DOF rigid transformation parameters.

The optimization goal during the 3D reconstruction process is to maximize the similarity between the biplanar DRR images and the lumbar decomposition images. Since the raw X-ray images have already been converted into DRR style, we chose a simple mean absolute error (MAE) loss as the image similarity metric. To emphasize the significance of the spinous and transverse processes during reconstruction, we customized a weighting matrix based on the landmark positions. The total weighted loss function can be expressed as:

$$Loss = \sum_{i=\{AP,LAT\}} \left| I_{pred}^{i} - I_{DRR}^{i} \right| \cdot W^{i} \tag{9}$$

Where $I_{pred}$ is the predicted lumbar decomposition image, $I_{DRR}$ is the DRR generated by the current model, and $W$ is the weight matrix of the same size as the image. We applied weighting to the positions of the transverse processes and spinous processes on $W$ using a gaussian point spread function:

$$W(x,y) = \sum_{i=1}^{15} \frac{1}{2\pi\sigma^2} e^{-\frac{(x-x_i)^2+(y-y_i)^2}{2\sigma^2}} + \lambda \tag{10}$$

Where $(x_i, y_i)$ are the landmark coordinates output by the network. $(x, y)$ are the points on the weight matrix. The generation radius for the weighting region is defined as $d$, which corresponds to the distance between the transverse process or spinous process and the pedicle landmark. Thus, $x \in [x_i - d, x_i + d], y \in [y_i - d, y_i + d]$. $\lambda$ denotes the base weight and $\sigma$ the standard deviation; through parameter tuning, they are set to 0.3 and 10, respectively. We focus on a pair of transverse processes and spinous processes for each vertebra, resulting in a total of $3\times 5 = 15$ landmarks were used to assign weights, as shown in Fig. 6.

Our objective function is non-convex and has multiple local extrema, which requires the optimizer to demonstrate high robustness against noise and irregular objective functions, as well as global convergence capability. Therefore, we selected the Bound Optimization BY Quadratic Approximation derivative-free optimization algorithm from the NLopt nonlinear optimization library for the registration process.

## 3 Experiments

### 3.1 A. Data Acquisition

The study was approved by the Institutional Review Board of Shanghai Tongji Hospital (protocol number: 2021-011-SK). We collected a dataset of 225 lumbar CT scans (United Imaging, uCT760, 120 kV, 512 × 512 resolution, 1.0 mm slice spacing). Subjects with abnormal osteophyte formation, obvious vertebral deformity, or fracture were excluded from the study. Furthermore, calibrated biplanar X-ray images (GOLDEN EYE, TAOiMAGE, Shanghai, China; image resolution 3072 × 3072 pixels) of the lumbar spine were acquired for 50 of the participants. During biplanar X-ray image acquisition, subjects were instructed to perform 6-DOF lumbar movements (flexion, extension, left/right bending, left/right rotation) at three different levels, and the corresponding biplanar X-ray images were captured simultaneously. During the experiment, all participants wore lead protective clothing in non-imaging areas to reduce unnecessary radiation exposure. Approximately 18 pairs of biplanar X-ray images were taken for each participant. According to the manufacturer's product manual, the effective dose for each pair of lumbar images under our typical test conditions (78 kV, 40 mA) was 0.16 mSv, which is less than 2.88 mSv per test, or about 14.4% of the average annual occupational exposure limit. Written informed consent was obtained from all participants before collecting any personal or health-related data. The demographics data of the subjects is shown in Table 2.

### 3.2 B. Data Labeling

The lumbar CT images were imported into the 3D visualization and modeling software Amira 6.7 (Thermo Fisher Scientific, Rockford, IL, United States), where an experienced orthopedic surgeon segmented the L1-L5 vertebrae and annotated seven bony landmarks (the centers of the superior and inferior endplates, the tips of the left and right transverse processes, the centers of the left and right pedicles, and the tip of the spinous process) on each lumbar vertebra.

To generate training labels for deep learning and evaluate the 3D reconstruction, the biplanar X-ray images were registered with the 3D CT-segmented L1-L5 vertebral models of the same participants. A custom registration program (MATLAB, R2023a, MathWorks, Natick, MA, United States) was used to adjust the 3D position and rotation of each vertebra until it matched the corresponding contours in the biplanar images. This technique has been validated as a gold standard, with a repeatability of less than 0.3 mm in translation and less than 0.7° in orientation when reproducing in-vivo human spine 6DOF kinematics [32]. Simulated projections were then conducted in the registered environment to generate DRRs of only the lumbar spine [33]. Additionally, the expert-annotated 3D vertebral landmark coordinates were projected onto the 2D images, and their 2D coordinates were calculated. The lumbar DRRs and landmark coordinates would serve as training labels for the X-ray image processing network.

### 3.3 C. Experimental settings

First, we constructed SSMs for each vertebra, using CT segmentation model from 175 subjects without biplanar X-ray images. To investigate the relationship between the number of SSM vertices, reconstruction accuracy, and reconstruction time, we remeshed the meshes in the dataset to specified vertex counts, creating SSMs with 8000, 4000, 2000, 1000, and 500 vertices, respectively.

Subsequently, the multi-task image processing network was trained. In a biplanar X-ray image dataset of 50 individuals, two subjects were diagnosed with L2-L3 vertebral fusion and lumbar sacralization, respectively. We designated their data as a special test set. From the remaining 48 individuals, we randomly selected 144 pairs of biplanar images from 8 subjects to form the test set. The remaining 720 pairs of biplanar images from 40 subjects were used for the training and validation set. We employed five-fold cross-validation to tune the hyperparameters. Prior to training, all data were resampled to a resolution of 512×512 pixels, with a corresponding pixel spacing of 0.834×0.834 mm. The training process was conducted on a workstation equipped with four NVIDIA GeForce RTX 3070 GPUs (with 24GB main memory). The optimizer used was AdamW, which separates weight decay from gradient updates, providing better regularization effects. The initial learning rate was set at 0.001, with a total of 300 training epochs, and the learning rate decayed to 0.9 times the original rate every 20 epochs. Additionally, we performed random flipping and rotation on the training set to increase data diversity, enhance the model's generalization capability, and improve robustness.

Ultimately, we performed 3D lumbar spine reconstruction on the test set for the image processing task, which consisted of 144 image pairs from 8 subjects. The five vertebrae are optimized simultaneously to improve reconstruction efficiency. This process was executed on a desktop computer equipped with an Intel(R) Core(TM) i7-14700KF CPU and an NVIDIA GeForce RTX 4060Ti GPU. A typical convergence curve for the 2D-3D registration process is shown in the Fig.

7. It is characterized by significant fluctuations during the first 200 iterations, followed by a pronounced decrease and gradual stabilization. Consequently, we employ a stopping condition based on a sliding-window convergence criterion, supplemented by a maximum iteration limit of 600. Let $f_k$ be the loss function value at the $k$ -th iteration. The process terminates when:

$$|f_{k-W} - f_k| < \epsilon_f \ AND \ std(\{f_{k-W}, f_{k-W+1}, \dots, f_k\}) < \sigma_f \tag{11}$$

where $W$ is the window length, set to 15. The parameters $\epsilon_f$ and $\sigma_f$ represent the convergence threshold and stability constraint, respectively, both of which are set to $10^{-3}$. This strategy ensures numerical stability while preventing premature termination of the registration procedure.

### 3.4 D. Ablation Experiments

In the ablation study, we kept all other settings constant, modifying only specific components to evaluate their impact on model performance. First, we evaluated the FFM in the image processing network to determine whether it could effectively relate features from the lumbar decomposition and landmark detection tasks. Then, we trained the lumbar spine segmentation and landmark detection tasks separately to determine whether the multi-task processing approach could improve model performance. Finally, we evaluated the landmark-weighted reconstruction loss in the 2D-3D registration process to determine whether it could improve the 3D reconstruction accuracy of the transverse and spinous processes.

### 3.5 E. Evaluation Metrics

The construction of the SSM was evaluated using two metrics, specificity and generalization. Specificity measures the model's ability to generate shapes similar to instances in the training set and is defined as the matching distance between uniformly randomly generated shapes and their closest matches in the training set. Generalization measures the model's ability to generate new instances, validated using the leave-one-out method, which is defined as the average error of all leave-one reconstructions [19].

We employed three performance metrics to quantitatively evaluate the lumbar decomposition task: Dice coefficient, peak signal-to-noise ratio (PSNR), and structural similarity index measure (SSIM). Dice is commonly used to assess image segmentation tasks, while PSNR and SSIM measure the results of image style transfer in DRR. In the landmark detection task, we utilized accuracy and success detection rate (SDR) to evaluate the model's predictions. Accuracy refers to the Euclidean distance between predicted landmark coordinates and ground truth. SDR indicates the success prediction rates for mean distances within 2mm, 4mm, and 6mm, respectively.

Since the 3D reconstruction process was conducted in a predefined biplanar X-ray projection environment, we were able to directly compare the reconstruction results to the gold standard (both pose and shape errors are reflected by the metric). We calculated the reconstruction accuracy (average symmetric point-to-surface distance) between the reconstructed model and the gold standard to measure the overall reconstruction error. Additionally, the Hausdorff distance was computed to quantify the maximum deviation. The lumbar lordosis angle, scoliosis angle, axial rotation angle, degree of spondylolisthesis, and intervertebral disc height were calculated based on the local coordinates of each vertebral model and the expert-annotated endplate positions. These results were then compared with the measurements taken by experts on the gold standard models to validate the clinical utility of the proposed method. These validations were performed on the models reconstructed using SSMs with 8000 vertices.

# 4 Results

## 4.1 A. Statistical Modeling

As the number of principal components increases, specificity gradually improves while generalization decreases, indicating that more principal components enable the model to better adapt to different scenarios, as shown in Fig. 8.

Fig. 9 illustrates the deformation of the statistical shape model of the L1 vertebra under three times the standard deviation of the first three principal components. The first principal component primarily represents the size of the vertebra, the second principal component mainly characterizes the width of the transverse processes, and the third principal component primarily reflects the thickness of the lamina, among other features.

## 4.2 B. X-ray Image Preprocessing

As illustrated in Fig. 10, the proposed network can accurately segment the lumbar spine from both AP and LAT X-ray images and convert them into a DRR style, while simultaneously achieving precise vertebral landmark detection. The quantitative metrics for vertebra separation are presented in Table 3, where our proposed method demonstrates excellent performance across Dice, PSNR, and SSIM scores. The Dice coefficients for AP and LAT images reached 0.995 and 0.993, respectively, indicating that the network can precisely segment the lumbar spine. The high PSNR and SSIM values also confirm the effectiveness of the style conversion. The accuracy and SDR for the landmark detection task are shown in Table 4 and Table 5. Detection accuracy varies among different landmarks, with the accuracy for endplate and pedicle centers being significantly higher than that for the transverse and spinous processes. Overall, both lumbar decomposition and landmark detection were more challenging on LAT images than on AP images. The results of the ablation study indicate that our method performed best, followed by the network without the FFM, with the single-task approach yielding the poorest performance.

The model's prediction results for vertebral fusion and lumbar sacralization are shown in Fig. 11 and Fig. 12, respectively. In the case of vertebral fusion, the model was able to successfully segment the fused vertebrae, even though no similar samples were present in the training set. Landmark detection for the endplate centers and transverse process tips of the fused vertebra was also successful. In the lumbar sacralization case, we had annotated the sacralized vertebra as L5, yet the model only outputted the decomposition results for the first four lumbar vertebrae. This occurred despite all annotated data containing five lumbar vertebrae. Concurrently, the model's landmark detection for the first four vertebrae was effective, with large detection errors appearing only on the sacralized L5.

## 4.3 C. Automatic 3D Reconstruction

All 144 cases in the test set successfully completed lumbar 3D reconstruction, with no instances of failure in pose initialization or convergence to a local optimum. As shown in Table 6, the proposed method achieved a mean 3D reconstruction error of $0.80 \pm 0.15$ mm and a Hausdorff distance of $5.56 \pm 1.29$ mm. Using the MAE as the loss function resulted in a mean reconstruction error of $0.93 \pm 0.34$ mm and a Hausdorff distance of $6.43 \pm 2.00$ mm. These results indicate that the proposed landmark-weighted reconstruction loss function yields lower reconstruction errors.

As illustrated in Fig. 13, we partitioned each vertebra into four regions—the vertebral body, transverse process, articular process, and spinous process—and evaluated the 3D reconstruction accuracy for each region using two different loss functions. The results, presented in Table 7 and Fig. 14, show that the reconstruction accuracy for the vertebral body was comparable between the two methods, with both achieving a mean error of approximately 0.62 mm. However, when using the landmark-weighted reconstruction loss function, the transverse, articular, and spinous processes were reconstructed more accurately, with all their mean errors below 1 mm. In contrast, using MAE as the loss function resulted in errors exceeding 1 mm for these parts. Fig. 15 illustrates the distribution of average errors between all reconstructed models and their corresponding ground truth. The reconstruction error was lowest in the vertebral body, while larger errors primarily occurred in the transverse processes, spinous processes, and articular facets. Compared to the simple MAE loss, our proposed landmark-weighted loss function achieves higher accuracy in reconstructing the complex structures of the vertebral arch.

The robustness of the proposed method was validated against significant postural changes in the lumbar spine. As shown in Table 8, the evaluation encompassed a wide range of motion, including large sagittal rotations (from -5.3°±13.9° in flexion to 42.3°±8.7° in extension), as well as major coronal (~±24–25°) and axial (~±8–9°) rotations. Despite these demanding conditions, the reconstruction accuracy was consistently high. The total error ranged narrowly from a minimum of 0.77 mm (in standing and flexion) to a maximum of 0.84 mm (in extension), demonstrating the method's stability.

As shown in the Table 9, the automated clinical parameter measurements from the reconstructed 3D lumbar models were robust, exhibiting small Mean Absolute Errors (MAEs). On the test set of 144 cases featuring diverse lumbar postures, the automated measurements achieved excellent reliability for the lordosis angle (ICC=0.97), scoliosis angle (ICC=0.96), degree of spondylolisthesis (ICC=0.93), and intervertebral disc height (ICC=0.90). The axial rotation angle demonstrated good agreement (ICC=0.70). These results confirm the accuracy and clinical utility of our method.

To evaluate the trade-off between accuracy and efficiency, we investigated the impact of the SSM vertex count on reconstruction error and time. As shown in Fig. 16, while increasing the number of vertices reduced the reconstruction error, it came at a significant computational cost. Notably, beyond 1000 vertices, the analysis revealed a point of diminishing returns: the accuracy gain was minimal (from 0.86 mm to 0.80 mm), whereas the reconstruction time increased substantially from 14.4 s to 36.5 s.

## 5 Discussion

In this study, we developed and validated a fully automatic and highly efficient method for sub-millimeter 3D reconstruction of the lumbar spine from biplanar X-ray images. Our main contribution lies in introducing for the first time a multi-task deep learning network dedicated to both lumbar decomposition and landmark detection. The lumbar decomposition effectively eliminates interference from surrounding tissues, simplifying the subsequent image similarity measurement in the registration process. Concurrently, the detected landmarks provide a reliable initial pose for the SSM registration, significantly accelerating convergence and preventing the process from settling into local optima. Furthermore, our proposed landmark-weighted reconstruction loss function successfully guides the model to focus on and enhance the reconstruction accuracy of complex posterior elements, such as the transverse and spinous processes. Finally, we validated the method's efficacy against a high-precision "gold standard," and the results demonstrate that our approach can complete a sub-millimeter accuracy lumbar reconstruction and reliable clinical parameter measurements within 20 seconds, achieving state-of-the-art performance in both efficiency and accuracy.

We observed that the landmark detection accuracy for the transverse and spinous processes was significantly lower than that for the endplates and pedicle centers. We attribute this primarily to the poor edge definition of these slender structures on X-ray images, which makes it difficult for the network to precisely capture their boundaries [34]. Additionally, both tasks' performance on lateral images was generally inferior to that on anteroposterior images, a finding consistent with the inherent high degree of structural overlap in the lateral view, as reported in other studies [35]. To address this, future research could explore cross-view feature fusion mechanisms, enabling the network to learn the geometric correspondence between landmarks in both views, thereby leveraging the clarity of the anteroposterior view to assist localization in the more ambiguous lateral view. Our ablation study clearly demonstrated the superiority of multi-task learning, indicating that combining lumbar decomposition and landmark detection encourages the network to learn the underlying spatial relationships between the two tasks, thus synergistically improving the prediction accuracy of both. Interestingly, our model correctly processed cases of vertebral fusion and, in a case of lumbar sacralization, accurately identified and segmented only the normal lumbar vertebrae. This suggests that our model has acquired a degree of anatomical "intelligence" and possesses the potential to distinguish between normal and abnormal vertebrae, paving the way for future applications on clinical pathological data.

Traditional 2D-3D registration methods typically employ mutual information or normalized cross-correlation as the optimization objective to handle the stylistic differences between DRRs and real X-rays, as well as noise and occlusions [19]. However, when target bone features are indistinct, such methods often require the aid of manual segmentation [26] or are limited to reconstructing only partial segments due to occlusion from other bones [16]. Aubert et al. proposed using a Generative Adversarial Network to translate X-ray images into a DRR-like style for registration [22,36]. However, the generative process compromises the network's ability to perform precise spinal segmentation, and the final reconstruction accuracy failed to reach the sub-millimeter level. In contrast, our study utilizes a high-precision multi-task network to directly extract clear vertebral regions and key geometric anchors from the original images. This "decomposition-localization" paradigm, combined with our custom-designed landmark-weighted loss function, is the key factor that enabled us to achieve sub-millimeter reconstruction accuracy for the first time. Furthermore, the detected landmarks facilitate the initial pose estimation, which enhances both the robustness and efficiency of the registration. It is also worth noting that reconstruction accuracy is closely tied to image resolution. The pixel width of our segmented images was 0.834 mm, and considering the magnification effect of cone-beam projection, the effective resolution is even higher. It can be said that our method has achieved a reconstruction accuracy approaching the physical limits of the imaging resolution.

Although our method achieves sub-millimeter accuracy, we believe that attaining CT-level precision requires the future optimization focus to shift from the registration algorithm to the shape model itself. We observed that after achieving high-precision image segmentation, the primary bottleneck limiting a more precise fit is the constrained deformability of the SSM. Due to data and computational resource limitations, we can only select a finite number of principal components as optimization parameters, which prevents the model from perfectly reconstructing individuals with rare vertebral morphologies. To overcome this bottleneck, future research could explore two promising directions. One approach involves fusion with end-to-end reconstruction. While recent end-to-end methods can theoretically handle arbitrary deformities, they lack the strong shape priors needed for high accuracy [37]. Integrating the SSM as a strong regularization term within an end-to-end framework may therefore be an effective way to balance flexibility and stability. An even more advanced direction lies in developing deep learning-driven shape priors. Drawing inspiration from existing research [38], a deep network could be trained to directly regress a deformation field containing shape statistics from X-ray images, thereby generating highly personalized 3D models. Such deep learning networks that fuse prior shape information represent a significant future direction for ultra-sparse 3D reconstruction.

This study has several limitations. First, our SSM was built using data that lacked deformed vertebrae, which restricts the model's ability to represent pathological morphologies. Second, our validation was primarily based on healthy or sub-healthy subjects, lacking systematic testing on complex pathological cases such as scoliosis or spondylolisthesis. Although we included images of subjects in different bending postures to simulate some deformation, the model's generalizability needs further validation on real patient data. Finally, our method relies on a calibrated biplanar X-ray system, which is not yet widely available in primary healthcare institutions. Therefore, exploring how to achieve robust 3D reconstruction from commonly available uncalibrated images is a critical issue to be addressed in the future.

## 6 Conclusion

This study proposed and validated a fully automatic, fast, and high-precision framework for 3D lumbar spine reconstruction. By performing vertebra separation and landmark detection on the original X-ray images and introducing landmark position weights during the 3D reconstruction process, we have, for the first time, elevated the accuracy of biplanar X-ray reconstruction to the sub-millimeter level. This method not only enhances the reconstruction accuracy of the vertebral arch but also demonstrates excellent performance across various degrees of lumbar rotation. With its advantages of being low-radiation and fully automatic, it can serve as a powerful tool to augment standard X-ray diagnostics, facilitating the personalized diagnosis of conditions such as spondylolisthesis and lumbar scoliosis.

**Data Availability Statement:** Data will be made available on request.

**Declaration of competing interest:** Tsung-Yuan Tsai got research support from TAOiMAGE Co., Ltd., China. Other authors declared that they had no conflicts of interest.

**Funding Statement:** This project was sponsored by National Natural Science Foundation of China [32371367], Pudong New Area Science & Technology and Development Fund [210H1147900], Shanghai Innovation Medical Device Application Demonstration Project [23SHS02800-06] and "Science and Technology Innovation Action Plan" of Science and Technology Commission of Shanghai Municipality [22S31906000, 23S31901000].

**Ethics Approval Statement:** The study was approved by the Institutional Review Board of Shanghai Tongji Hospital (protocol number: 2021-011-SK).

**Patient Consent Statement:** Informed consent was obtained from all subjects involved in the study.

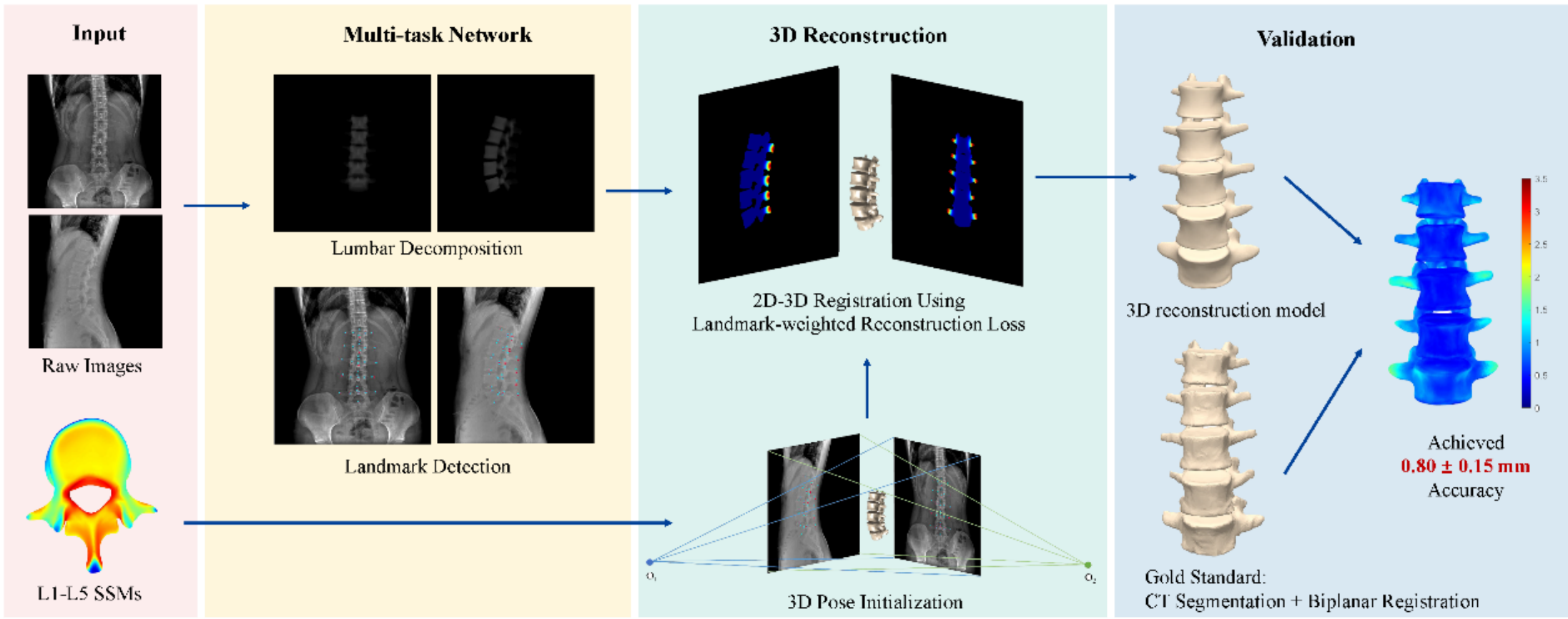

**Fig. 1.** Overall framework for the proposed 3D reconstruction of the lumbar spine from biplanar X-ray images.

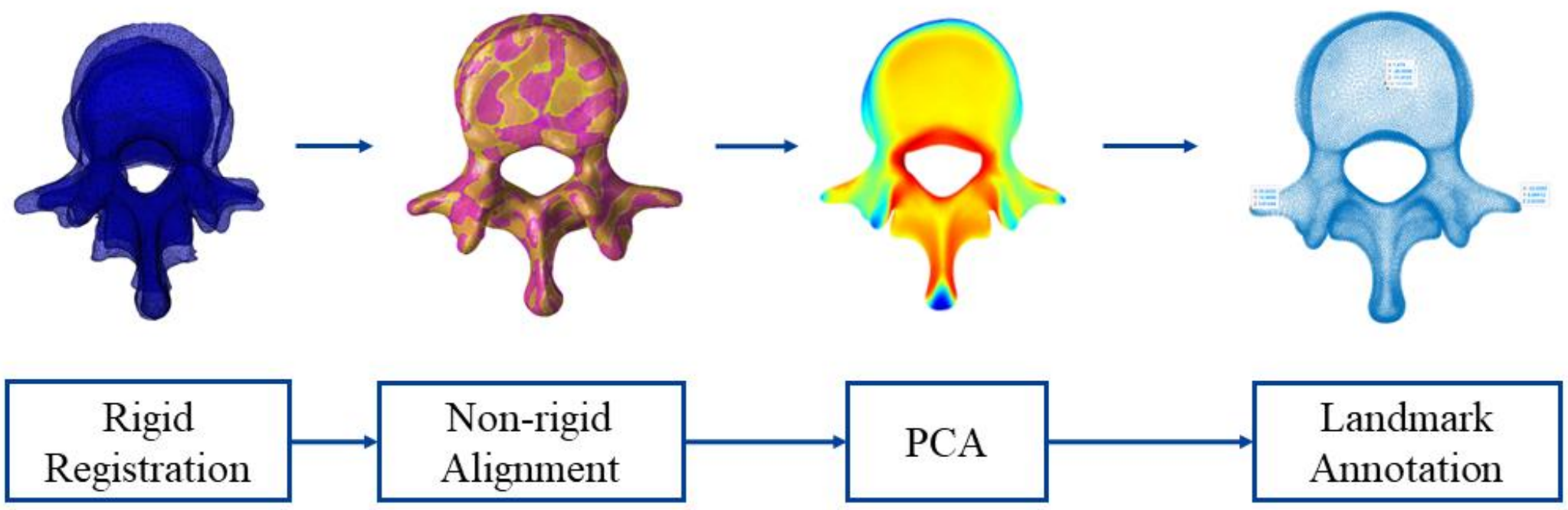

**Fig. 2.** Workflow for Establishing a SSM of Vertebrae.

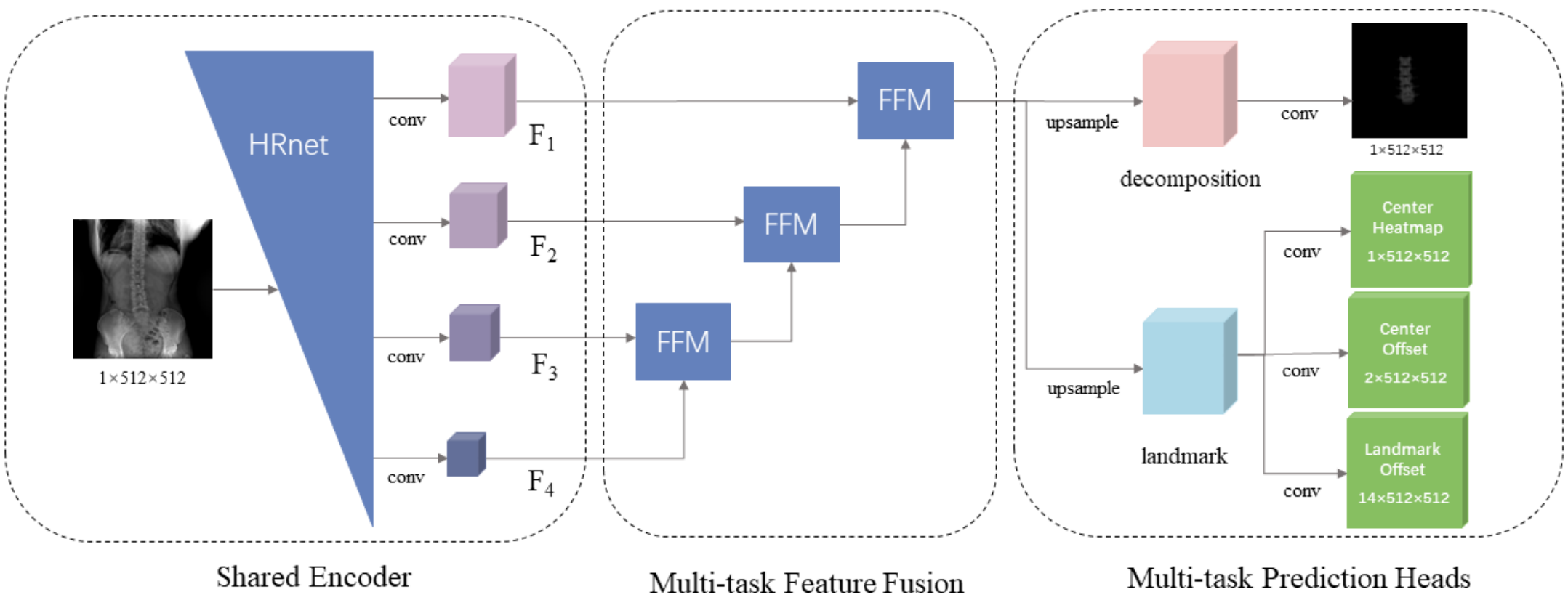

**Fig. 3.** Architecture of the multi-task processing network for lumbar decomposition and vertebral landmark detection.

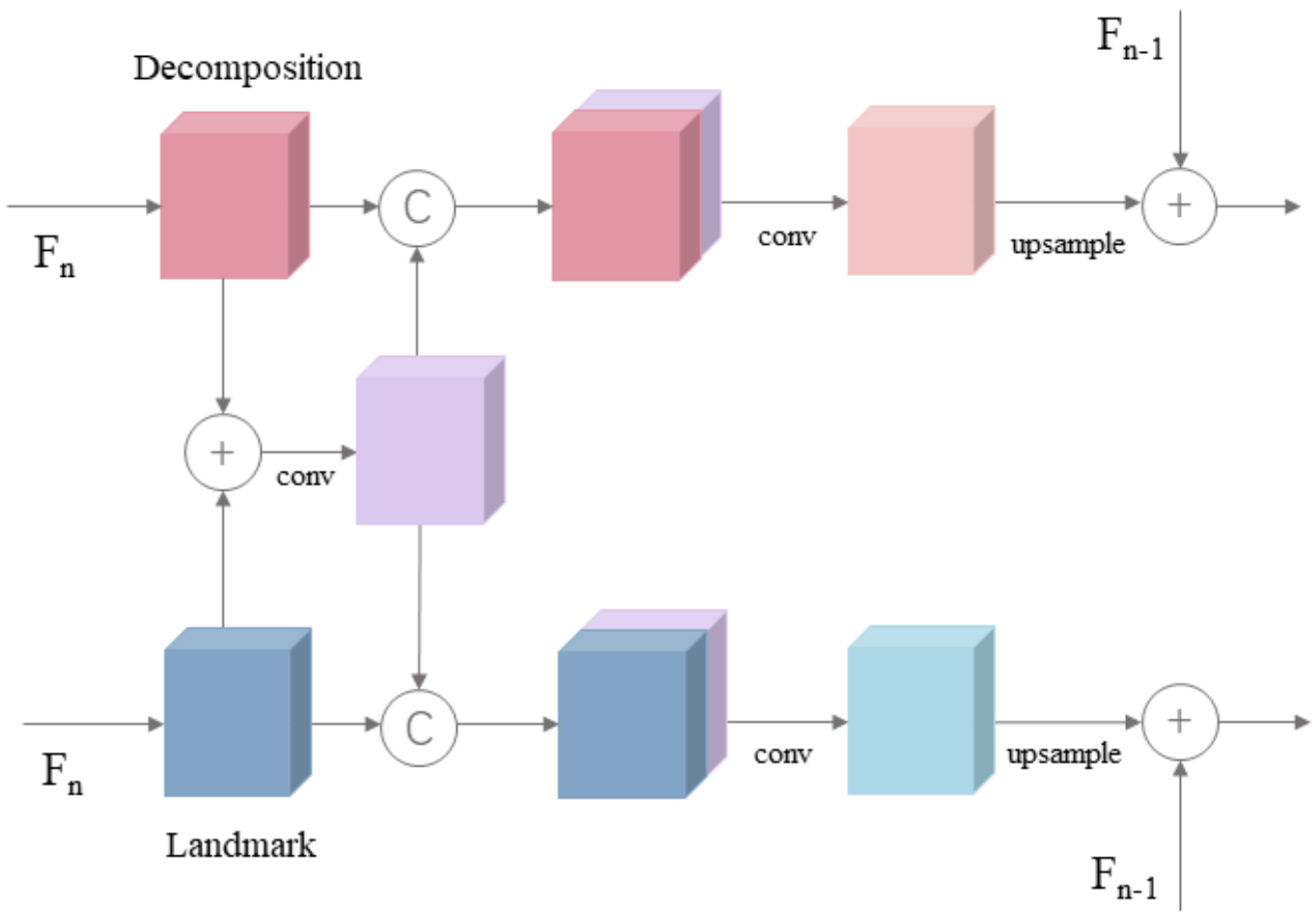

**Fig. 4.** A block diagram of the two-task feature fusion module.

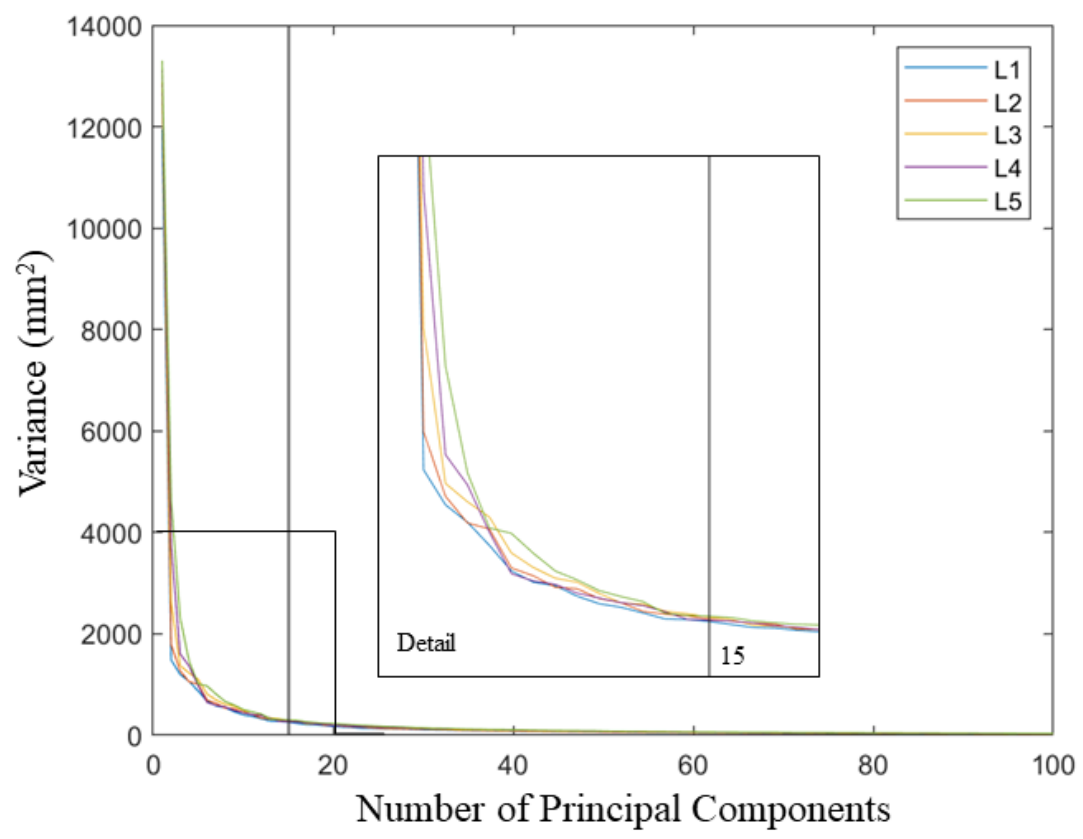

**Fig. 5.** Scree plot of the SSM for the L1-L5 vertebrae. The inflection point is at the 15th principal component.

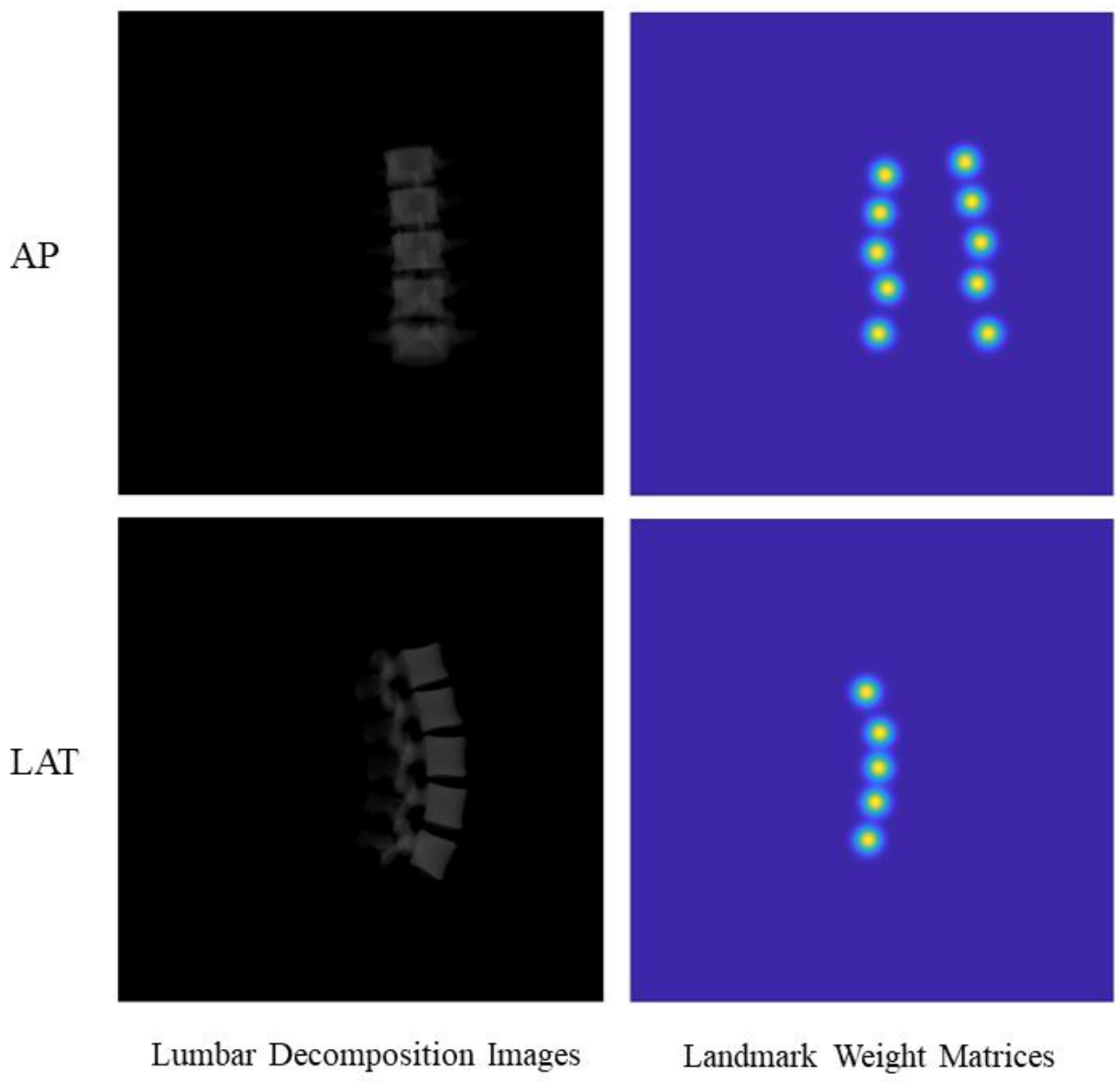

**Fig. 6.** The AP and LAT lumbar decomposition images and their corresponding landmark weight matrices.

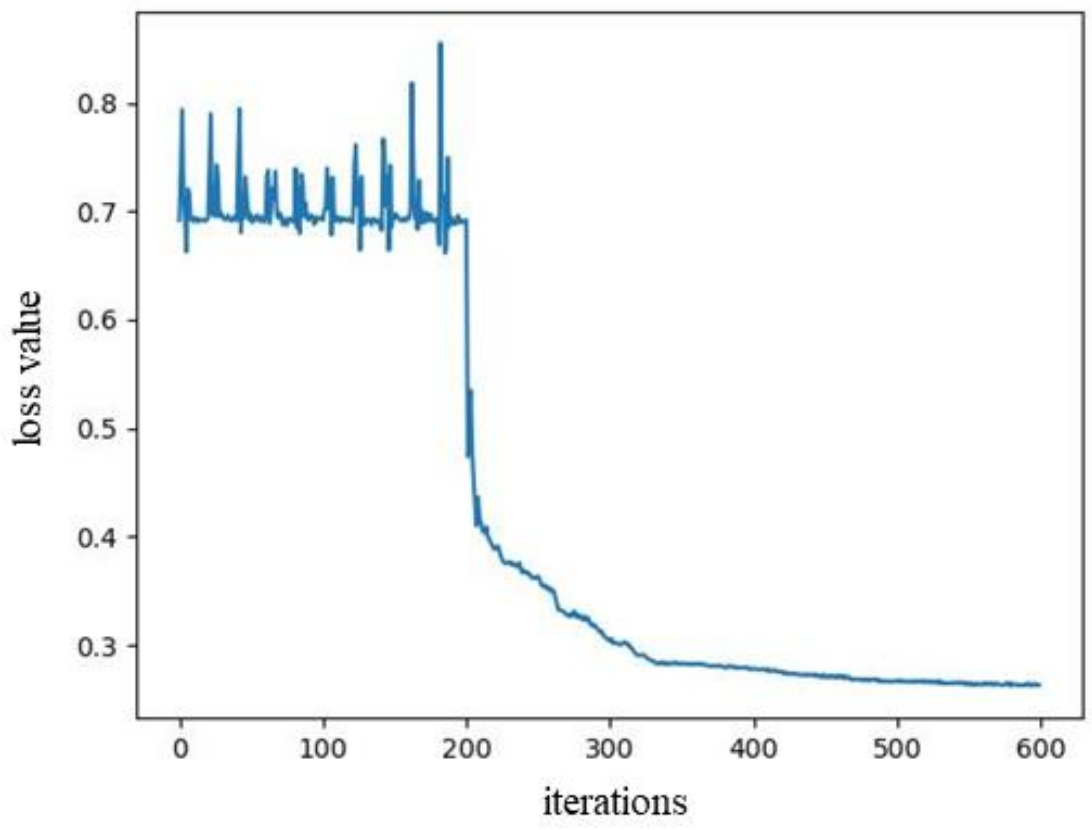

**Fig. 7.** A typical convergence curve for the 2D-3D registration process.

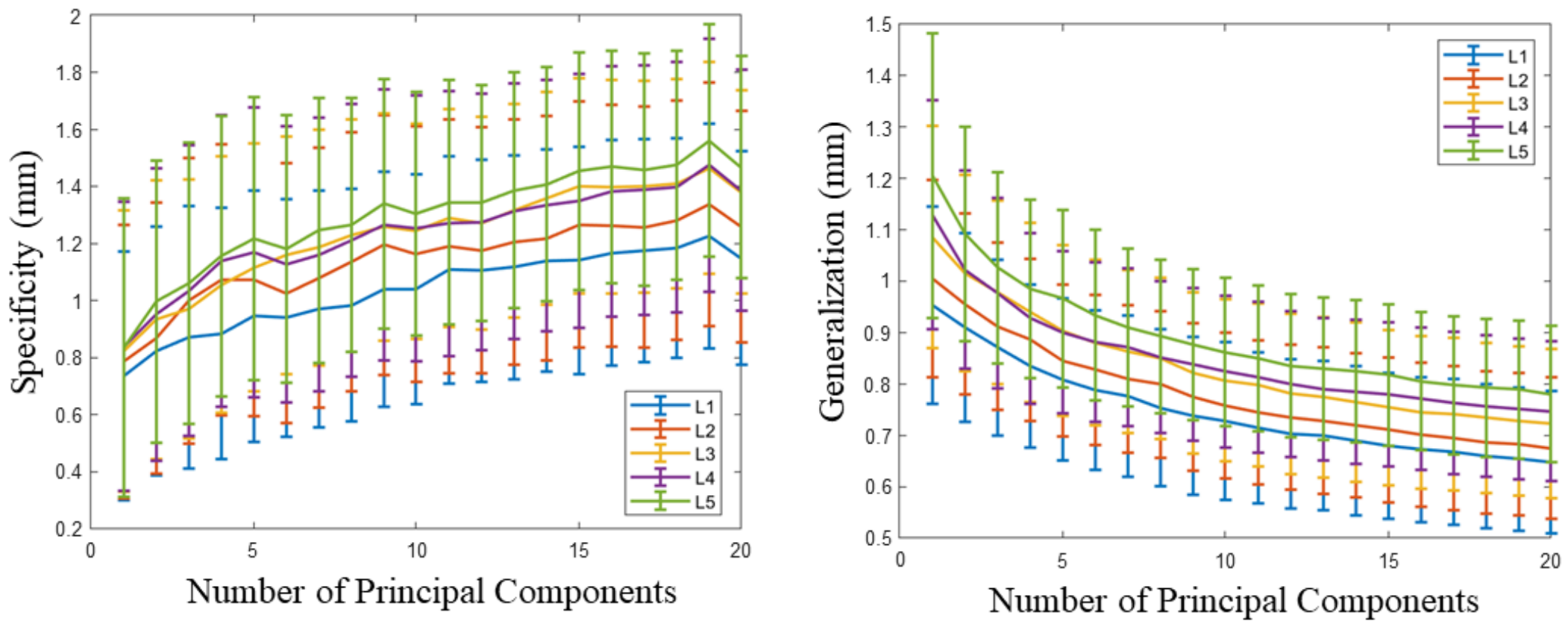

**Fig. 8.** The functional relationship between the specificity (left) and generalization (right) of the L1-L5 vertebral SSM and the number of principal components.

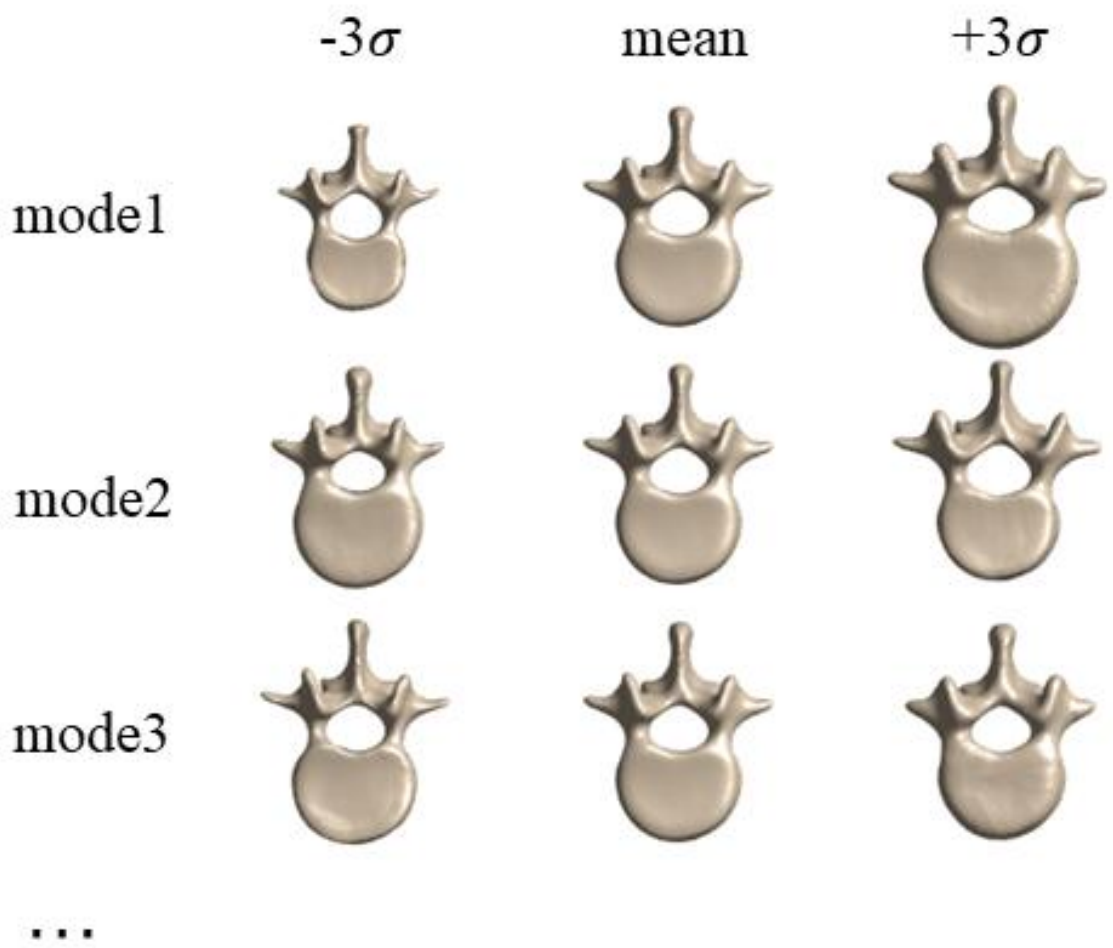

**Fig. 9.** The average of the SSM of the L1 vertebra from a top view and the first three variation modes, described using the average ± three times the standard deviation.

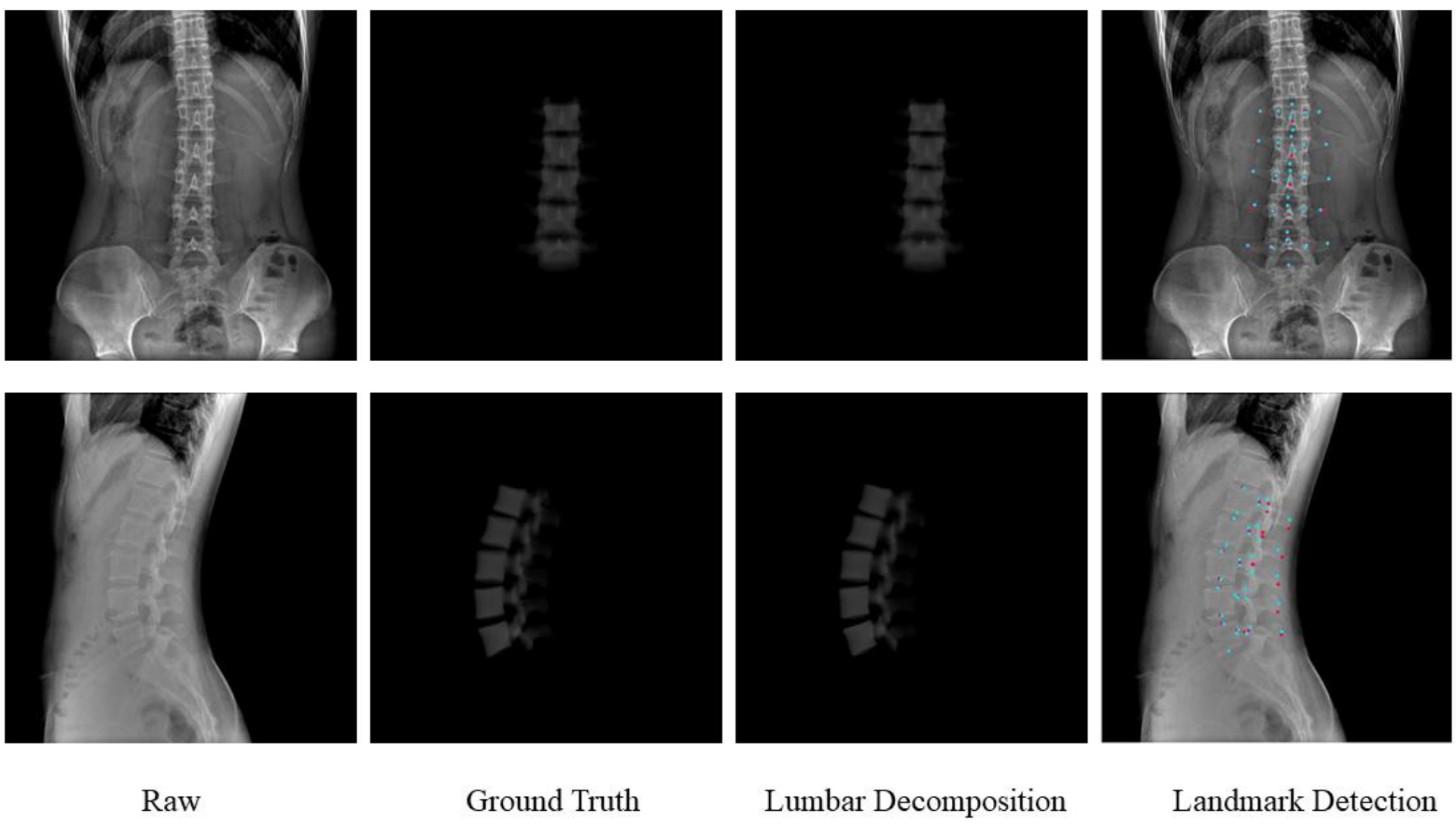

**Fig. 10.** Lumbar decomposition and landmark detection results output by the network. In the landmark detection result, blue marked points represent the ground truth, while red marked points indicate the predicted results.

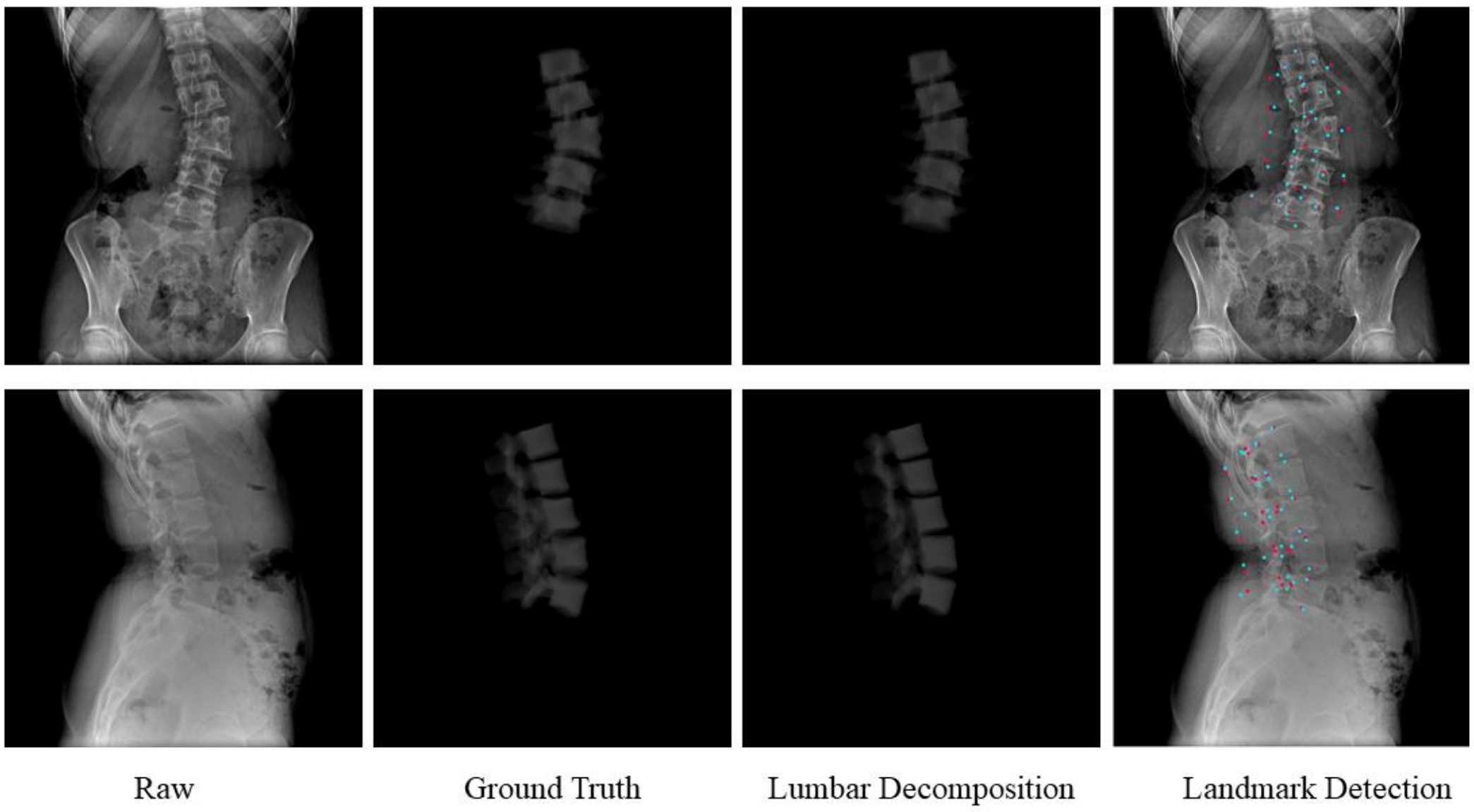

**Fig. 11.** Model prediction results of biplanar X-ray images in a patient with vertebral fusion.

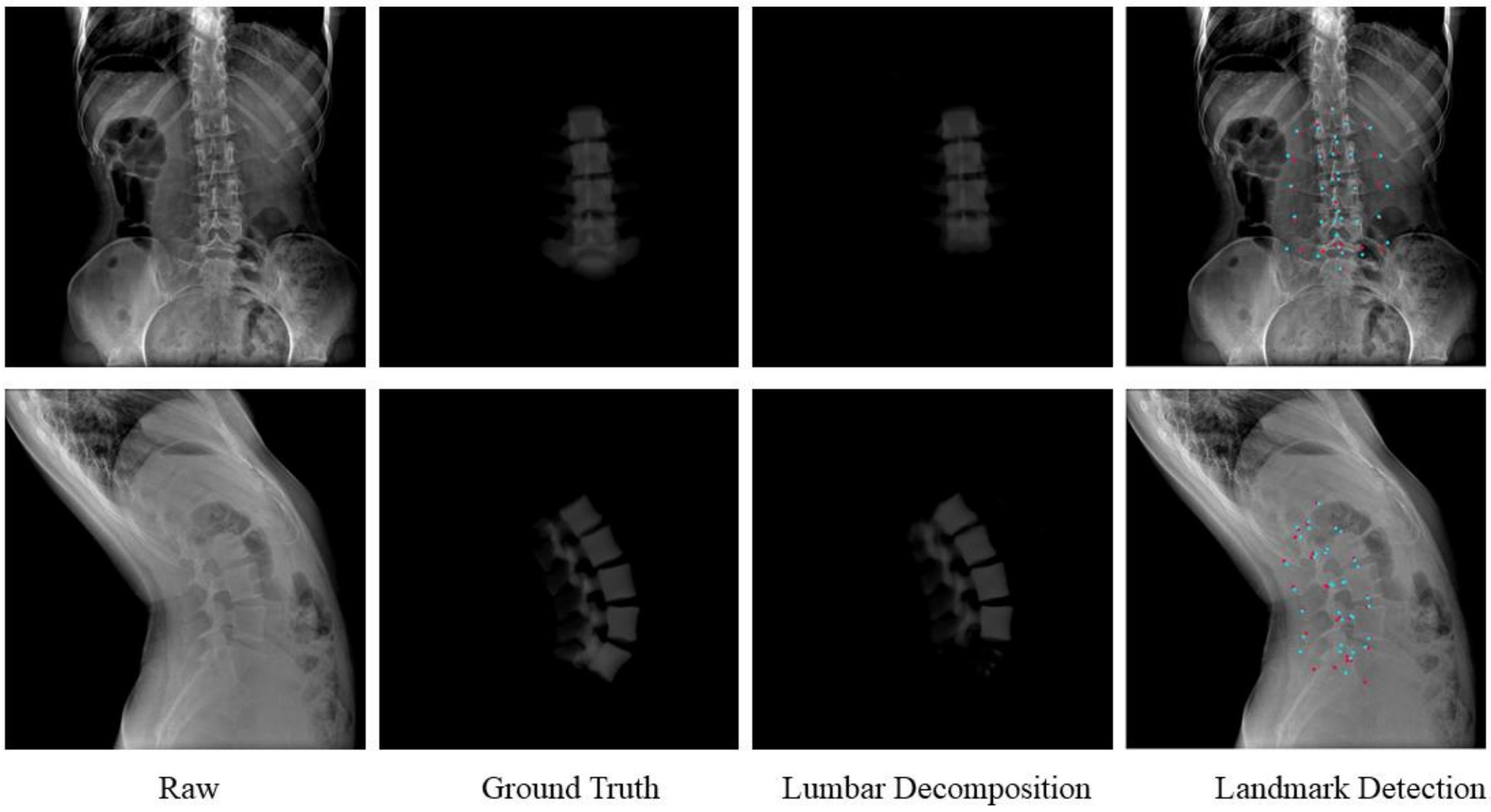

**Fig. 12.** Model prediction results of biplanar X-ray images in a patient with lumbar sacralization.

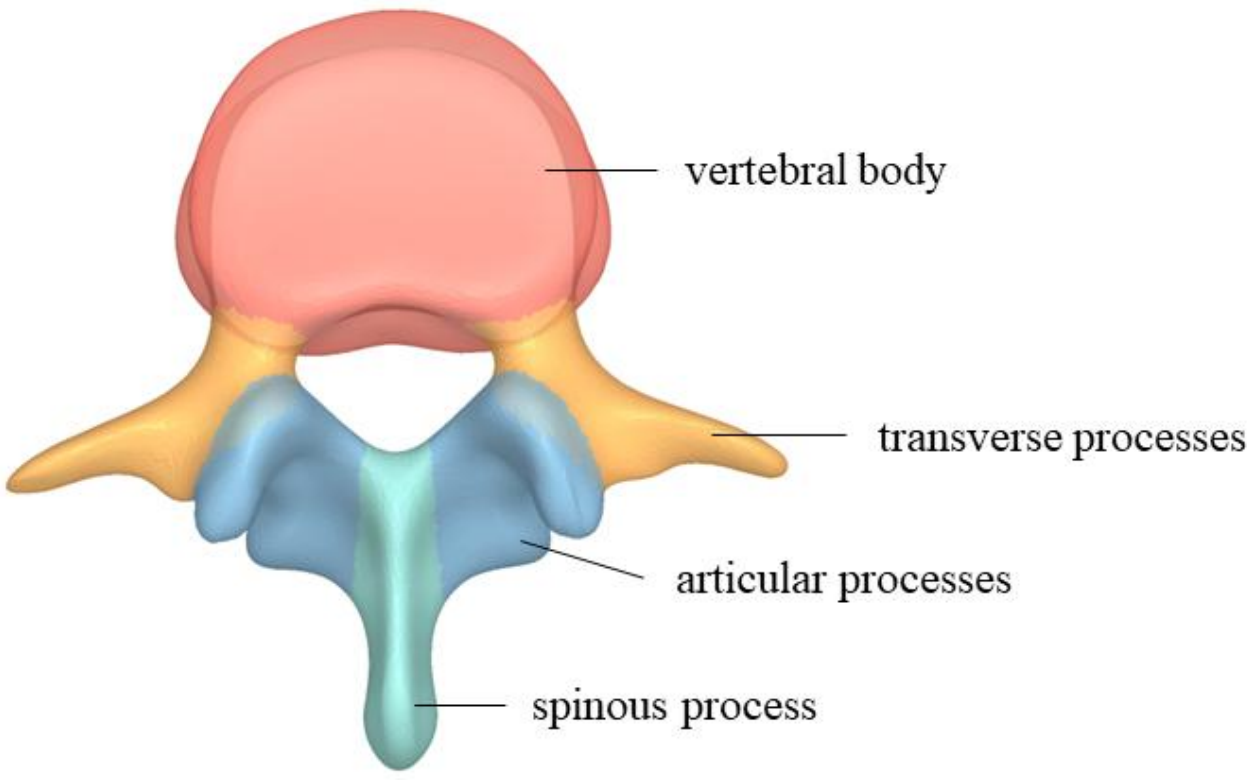

**Fig. 13.** The four regions of a vertebra: vertebral body (red), transverse processes (yellow), articular processes (blue), and spinous process (green).

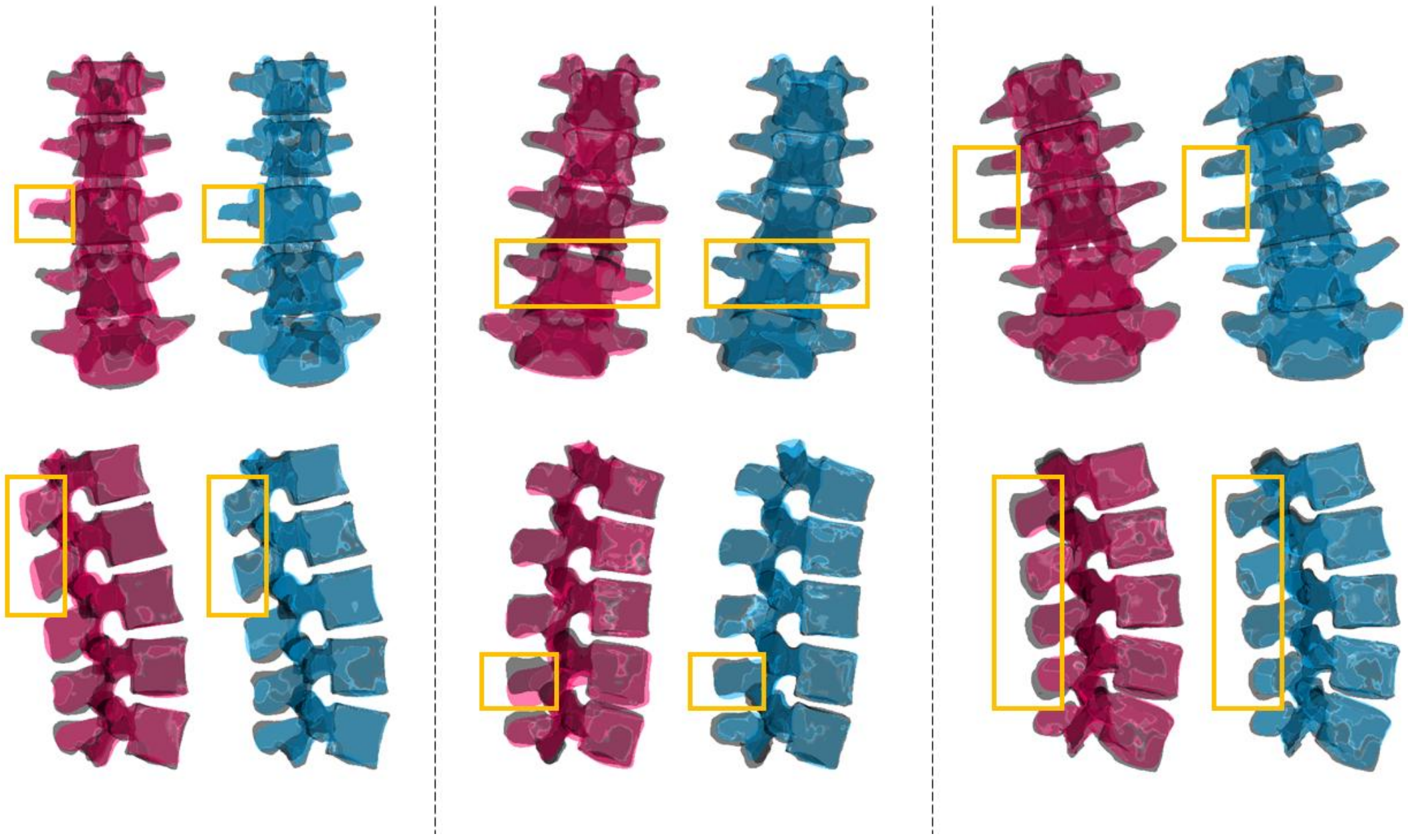

**Fig. 14.** 3D reconstruction results for three subjects. The black mesh represents the ground truth, the red mesh represents the reconstruction results using simple MAE loss function, and the blue mesh represents the reconstruction results using the landmark-weighted loss function.

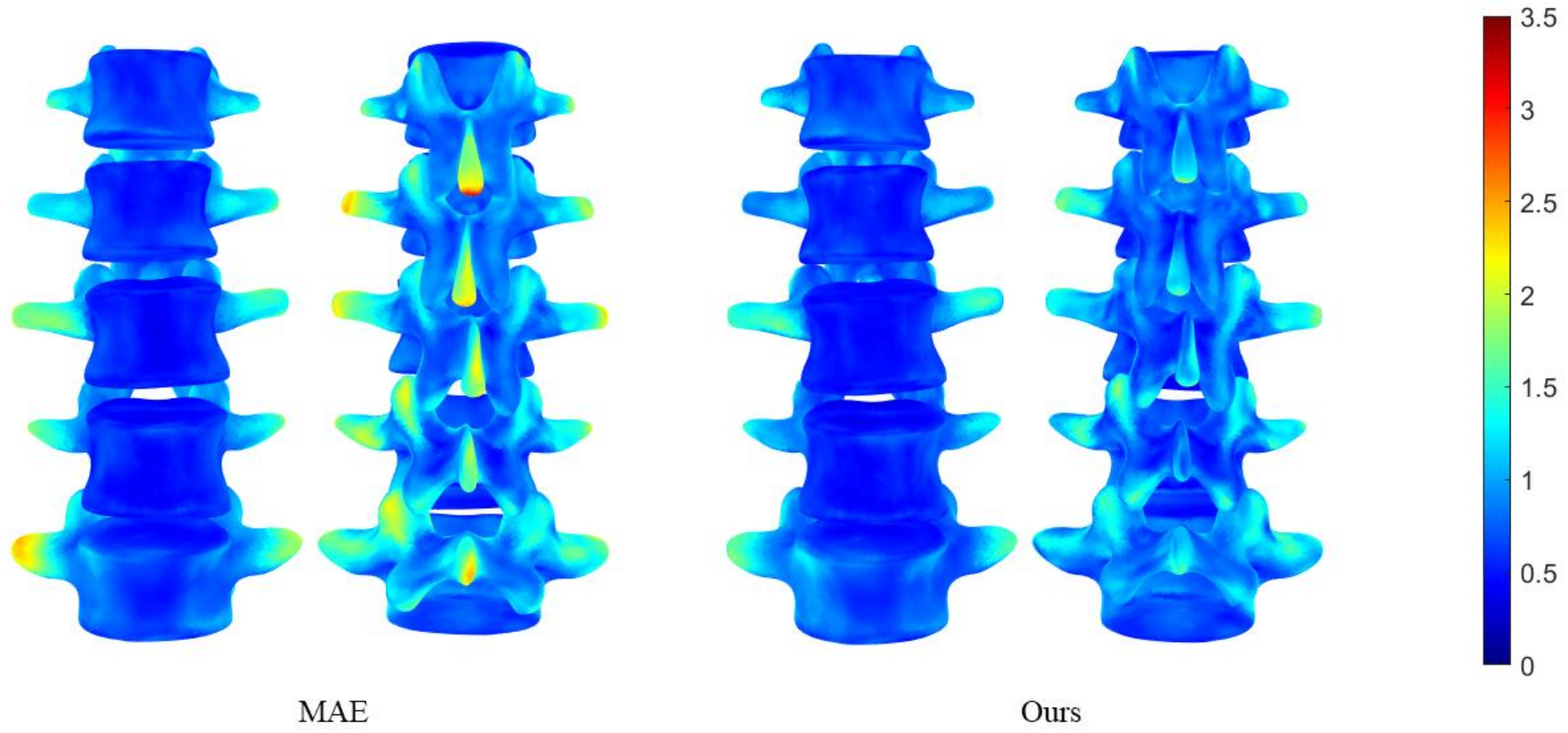

**Fig. 15.** Heatmap of average error distribution for 3D reconstruction. The color bar on the right represents the average error (mm) for different anatomical regions.

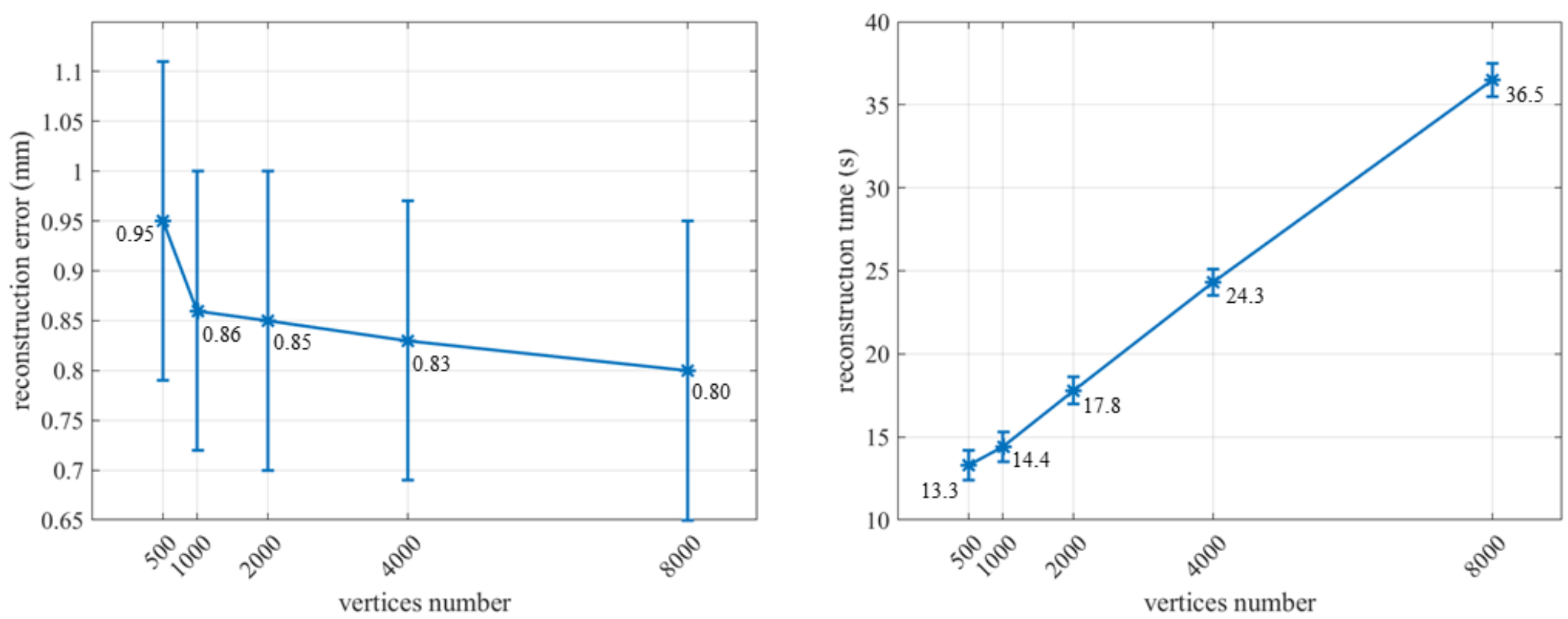

**Fig. 16.** Relationship between SSM vertex count and reconstruction error (left) and reconstruction time (right).

**Table 1.** Performance metrics of variable model-based lumbar spine 3D reconstruction methods.

| Study | Accuracy (mm) | Time Taken | Automation Level | Validation Standard |
|---|---|---|---|---|
| [16] | 1.2 | 3 minutes | Semi-automatic | CT/MRI segmentation |
| [26] | 1.7–2.0 | - | Semi-automatic | CT/MRI segmentation |
| [39] | 0.93–1.34 | 4.3 hours | Semi-automatic | CT/MRI segmentation |
| [18] | 1.9–2.2 | 34 seconds | Fully automatic | Manual deformation |
| [22] | 1.3-1.6 | <1minute | Fully automatic | Manual deformation |
| [36] | 1.0-1.6 | - | Fully automatic | CT/X-ray alignment |

**Table 2.** Subject demographics data.

| Variables | CT (N=175) | CT+ X-ray (N=50) |
|---|---|---|
| Gender(M:F) | 82:93 | 18:32 |
| Age(year) | 37.6±10.7 | 23.7±2.1 |
| Height(m) | 1.67±0.06 | 1.68±0.06 |
| Weight(kg) | 64.3±11.6 | 61.4±10.7 |
| BMI(kg/m$^2$) | 23.6±3.7 | 21.8±2.5 |

**Table 3.** Lumbar decomposition results.

| | | Dice | PSNR | SSIM |
|---|---|---|---|---|
| AP | **Ours** | **0.995** | **45.8** | **0.992** |
| | WO-FFM | 0.994 | 45.5 | 0.991 |
| | Single-task | 0.993 | 45.3 | 0.990 |
| LAT | **Ours** | **0.993** | **43.0** | **0.988** |
| | WO-FFM | 0.992 | 42.4 | 0.987 |
| | Single-task | 0.992 | 42.3 | 0.986 |

WO-FFM represents the multi-task network without the FFM module in our framework, while Single-task refers to the network using HRNet for single-task prediction.

**Table 4.** Accuracy of the landmark detection task (mm).

| | | Endplate center | Transverse process | Pedicle | Spinous process | Total |
|---|---|---|---|---|---|---|
| AP | **Ours** | **1.52** | **3.03** | **1.43** | **4.29** | **2.32** |
| | WO-FFM | 1.59 | 3.45 | 1.57 | 4.34 | 2.51 |
| | Single-task | 1.63 | 3.77 | 1.68 | 4.61 | 2.68 |
| LAT | **Ours** | **1.63** | **6.49** | **2.78** | **5.66** | **3.92** |
| | WO-FFM | 1.72 | 7.53 | 3.16 | 5.78 | 4.37 |
| | Single-task | 1.96 | 8.76 | 3.83 | 6.38 | 5.07 |

**Table 5.** SDR of the landmark detection task.

| | | endplate center | | | transverse process | | | pedicle | | | spinous process | | |
|---|---|---|---|---|---|---|---|---|---|---|---|---|---|
| | | 2mm | 4mm | 6mm | 2mm | 4mm | 6mm | 2mm | 4mm | 6mm | 2mm | 4mm | 6mm |
| AP | **Ours** | **75.2** | **98.7** | **99.9** | **30.3** | **75.4** | **94.3** | **78.1** | **98.9** | **99.9** | **21.7** | **53.8** | **78.9** |
| | WO-FFM | 73.7 | 98.5 | 99.8 | 25.2 | 67.3 | 91.1 | 73.4 | 98.1 | 99.8 | 20.2 | 51.2 | 75.5 |
| | Single-task | 72.6 | 97.6 | 99.6 | 21.2 | 62.6 | 87.4 | 71.5 | 97.9 | 99.6 | 19.3 | 48.2 | 72.5 |
| LAT | **Ours** | **69.8** | **98.5** | **99.8** | **8.6** | **30.6** | **52.2** | **40.5** | **78.9** | **94.7** | **13.1** | **37.6** | **59.0** |
| | WO-FFM | 67.1 | 97.3 | 99.6 | 7.9 | 24.5 | 42.6 | 30.6 | 72.8 | 92.0 | 10.4 | 34.9 | 59.4 |
| | Single-task | 67.1 | 97.3 | 99.2 | 5.3 | 18.7 | 36.1 | 25.1 | 63.7 | 86.9 | 8.3 | 30.0 | 55.2 |

**Table 6.** The reconstruction accuracy (mm) and Hausdorff distance (mm) for 3D Reconstruction.

| | | L1 | L2 | L3 | L4 | L5 | total |
|---|---|---|---|---|---|---|---|
| Accuracy | MAE | 0.90±0.32 | 0.91±0.30 | 0.91±0.36 | 0.92±0.38 | 1.01±0.34 | 0.93±0.34 |
| | **Ours** | **0.79±0.17** | **0.77±0.12** | **0.79±0.14** | **0.78±0.13** | **0.88±0.16** | **0.80±0.15** |
| Hausdorff | MAE | 6.15±1.68 | 6.18±1.94 | 6.08±2.04 | 6.74±2.27 | 6.99±1.90 | 6.43±2.00 |
| | **Ours** | **5.56±1.16** | **5.00±1.07** | **5.29±1.34** | **6.01±1.41** | **5.94±1.19** | **5.56±1.29** |

MAE represents the results obtained using simple average absolute error loss function.

**Table 7.** The reconstruction accuracy (mm) for different vertebral regions.

| | | L1 | L2 | L3 | L4 | L5 | total |
|---|---|---|---|---|---|---|---|
| Vertebral body | MAE | **0.64±0.23** | 0.61±0.18 | 0.56±0.17 | 0.60±0.18 | 0.72±0.21 | 0.62±0.20 |
| | **Ours** | 0.65±0.22 | **0.58±0.12** | **0.55±0.11** | **0.58±0.11** | **0.72±0.19** | **0.62±0.17** |
| Transverse process | MAE | 0.97±0.43 | 1.19±0.46 | 1.29±0.66 | 1.16±0.51 | 1.17±0.51 | 1.16±0.53 |
| | **Ours** | **0.84±0.19** | **0.98±0.25** | **1.08±0.37** | **0.99±0.23** | **0.96±0.22** | **0.97±0.27** |
| Articular process | MAE | 1.01±0.36 | 0.99±0.33 | 1.00±0.39 | 1.06±0.42 | 1.17±0.39 | 1.04±0.39 |
| | **Ours** | **0.89±0.20** | **0.88±0.21** | **0.89±0.25** | **0.92±0.25** | **0.98±0.22** | **0.91±0.23** |
| Spinous process | MAE | 1.42±0.76 | 1.24±0.66 | 1.24±0.85 | 1.32±1.04 | 1.49±0.98 | 1.34±0.87 |
| | **Ours** | **0.96±0.34** | **0.81±0.20** | **0.87±0.22** | **0.85±0.23** | **1.08±0.35** | **0.91±0.29** |

**Table 8.** Rotational angles of the lumbar and the corresponding 3D reconstruction accuracy across different postures.

| | Lumbar rotational angles (°) | | | Reconstruction accuracy (mm) | | | | | |
|---|---|---|---|---|---|---|---|---|---|
| | Lordosis | Scoliosis | Axial Rotation | L1 | L2 | L3 | L4 | L5 | TOTAL |
| Stand | 26.8±11.3 | 1.2±5.7 | 0.5±3.4 | 0.74±0.08 | 0.73±0.14 | 0.78±0.18 | 0.74±0.12 | 0.84±0.12 | 0.77±0.13 |
| Extension | 42.3±8.7 | 0.8±5.3 | 0.7±4.6 | 0.84±0.06 | 0.80±0.14 | 0.81±0.15 | 0.81±0.14 | 0.92±0.12 | 0.84±0.13 |
| Flexion | -5.3±13.9 | 0.7±7.5 | 0.9±6.2 | 0.79±0.07 | 0.73±0.07 | 0.74±0.12 | 0.77±0.12 | 0.82±0.09 | 0.77±0.10 |
| Right bend | 30.7±10.8 | 24.6±11.7 | 3.4±5.7 | 0.76±0.09 | 0.79±0.11 | 0.76±0.16 | 0.81±0.12 | 0.86±0.13 | 0.80±0.12 |
| Left bend | 31.1±11.6 | - 23.9±12.4 | -3.6±4.9 | 0.78±0.15 | 0.74±0.11 | 0.79±0.13 | 0.76±0.12 | 0.86±0.20 | 0.79±0.14 |
| Right twist | 29.5±10.2 | -5.3±5.1 | 8.7±1.6 | 0.76±0.10 | 0.78±0.13 | 0.83±0.12 | 0.80±0.13 | 0.92±0.13 | 0.82±0.13 |
| Left twist | 28.7±106 | 4.9± | -8.3±1.4 | 0.79±0.09 | 0.78±0.12 | 0.81±0.13 | 0.79±0.12 | 0.92±0.08 | 0.82±0.12 |

Positive values indicate extension in the sagittal plane, right bending in the coronal plane, and right twist in the transverse plane.

**Table 9** Comparison between automated measurements of lumbar clinical parameters and the gold standard.

| Parameter | Gold Standard | Auto Measured | MAD±SD | ICC[95%CI] |
| --- | --- | --- | --- | --- |
| Lordosis Angle (°) | 32.7±11.0 | 32.2±10.3 | 2.1±1.6 | 0.97[0.95-0.99] |
| Scoliosis Angle (°) | 1.6±8.7 | 1.7±8.8 | 1.5±1.4 | 0.96[0.94-0.98] |
| Axial Rotation Angle (°) | 0.8±3.2 | 0.7±3.6 | 1.7±1.6 | 0.70[0.62-0.78] |
| Degree of Spondylolisthesis (%) | 2.1±8.1 | 2.4±9.6 | 2.6±2.0 | 0.93[0.92-0.94] |
| Intervertebral Disc Height (mm) | 7.7±1.2 | 8.2±1.1 | 0.6±0.5 | 0.90[0.88-0.92] |